\documentclass[a4paper,fleqn]{cas-dc}

\usepackage{float}
\usepackage{graphicx}
\usepackage[numbers,sort&compress]{natbib}
\usepackage{placeins}
\usepackage{xcolor}
\usepackage{etoolbox} 
\usepackage{xspace}   
\usepackage{capt-of}
\usepackage{cuted}
\usepackage{xcolor}
\usepackage{booktabs}
\usepackage{multirow}
\usepackage{comment}

\definecolor{citeblue}{RGB}{0,120,170}

\def\tsc#1{\csdef{#1}{\textsc{\lowercase{#1}}\xspace}}
\tsc{WGM}
\tsc{QE}
\tsc{EP}
\tsc{PMS}
\tsc{BEC}
\tsc{DE}

\begin{document}
\hypersetup{
    colorlinks=true,
    citecolor=citeblue,
    linkcolor=citeblue,
    urlcolor=citeblue
}
\let\WriteBookmarks\relax
\def\floatpagepagefraction{1}
\def\textpagefraction{.001}

\shorttitle{pristine and Co-doped HfSSe monolayers as the anode materials}

\title [mode = title]{Enhancement of Li and Na Storage in Janus HfSSe Monolayers by Co-Induced Electron-Deficient Chalcogen Sites: A First-Principles Study}



%
\shorttitle{Janus HfSSe and Co-Doped HfSSe as Anode Materials}

   \shortauthors{M. M. Hasan et al.}


\author[1]{Md. Muktadir Hasan}[orcid=0009-0007-5899-4481]

\credit{Writing -- original draft, Visualization, Software, Methodology, Investigation, Conceptualization}
\author[1]{Md. Mafizul Islam}[orcid=0009-0003-6160-2726]

  \credit{Investigation}
\author[1]{Ahmed Zubair}[orcid=0000-0002-1833-2244]
\cormark[1]

\credit{Writing -- review \& editing, Validation, Supervision, Resources, Methodology, Investigation, Conceptualization}

\ead{ahmedzubair@eee.buet.ac.bd} 

\affiliation[1]{
    organization={Department of Electrical and Electronic Engineering, Bangladesh University of Engineering and Technology},
    city={Dhaka},
    country={Bangladesh}
}

\cortext[cor1]{Corresponding author}

\begin{abstract}
The growing demand for high-capacity, fast-charging, and durable rechargeable batteries has intensified the search for advanced electrode materials, with two-dimensional materials offering large surface-to-mass ratios and short ion-diffusion pathways. Here, pristine and Co-doped Janus HfSSe monolayers were investigated as anode materials for Li-ion and Na-ion batteries using first-principles density functional theory. Pristine HfSSe was found to be dynamically stable and exhibited a 0.85~eV band gap, while Co substitution shifted the Fermi level toward the valence-band region. Adsorption of the first Li or Na layer induced metallic behavior, thereby improving charge transport. Co doping strengthened alkali-metal binding, with the most favorable Li and Na adsorption energies changing from $-2.6125$ and $-1.9800$~eV to $-3.7405$ and $-3.0529$~eV, respectively. Co doping created electron-deficient sites that enhanced Li/Na binding. Co-doped HfSSe accommodated up to four Li layers and three Na layers, increasing the theoretical capacities from 555.43 to 776.19~mAh~g$^{-1}$ for Li and from 370.29 to 582.14~mAh~g$^{-1}$ for Na. The average open-circuit voltages were 0.62~V for Li and 0.29--0.30~V for Na, with minimum migration barriers of 0.2153 and 0.1044~eV, respectively. These results identified Co-doped Janus HfSSe as a promising anode material for Li-ion and Na-ion batteries.
\end{abstract}


\begin{highlights}

\item Co substitution creates electron-deficient chalcogen sites in Janus HfSSe.

\item These empty states accept the Li/Na electron, strengthening binding by $\sim$1.1~eV.

\item Co doping raises capacity to 776 (Li) and 582 (Na) mAh~g$^{-1}$ via multilayer storage.

\item Low average voltages of 0.62~V (Li) and 0.30~V (Na) favor anode operation.

\item Low barriers of 0.22~eV (Li) and 0.10~eV (Na) indicate fast ion diffusion.

\end{highlights}


\maketitle

\section{Introduction}
The increasing demand for portable electronics, electric vehicles,
grid-level storage, and renewable-energy integration has accelerated the
search for rechargeable batteries with high energy density, long cycle
life, fast charge--discharge capability, and improved safety. Among the
available electrochemical energy-storage technologies, lithium-ion
batteries (LIBs) have become the dominant rechargeable battery technology
because of their high energy and power densities, long cycle life, and
low self-discharge rate \cite{Khan2023LIB}. Their role has further
expanded in electric transportation, where high-energy-density and
fast-charging batteries are essential for widespread EV adoption
\cite{Zeng2023XFC}. Li-ion battery energy-storage systems have also been
increasingly deployed for grid-connected renewable-energy applications
\cite{Zheng2024GridBESS}. However, concerns regarding lithium supply,
price volatility, and dependence on critical-material supply chains have
stimulated increasing interest in alternative battery chemistries,
particularly sodium-ion batteries (SIBs) \cite{Yao2025Sodium}. SIBs are
particularly attractive for large-scale energy storage because of the
high abundance and broad availability of sodium resources and their
potential for lower-cost energy storage \cite{Wu2024Sodium}. Despite these advantages, battery performance strongly depends on the properties of the electrode materials, particularly the anode. Important anode-related characteristics such as ion adsorption, electronic conductivity, diffusion kinetics, operating voltage, storage capacity, and structural stability directly influence electrochemical performance \cite{obrovac2014alloy}. Therefore, rational design and modification of anode materials remain essential for achieving high capacity, rapid ion transport, improved stability, and fast-charging performance \cite{chen2020tmd}.

The conventional commercial anode material for LIBs is graphite, which
has a theoretical specific capacity of 372~mAh~g$^{-1}$~\cite{gra372}, corresponding
to the fully lithiated LiC$_6$ composition.
Experimentally, pristine commercial anode-grade graphite has delivered
a reversible specific capacity of approximately 356~mAh~g$^{-1}$ at
1C~\cite{Olutogun2024Graphite}. Sluggish ion-diffusion kinetics under high-rate operation limits the power density and fast-charging capability of graphite anodes \cite{goriparti2014review}. Moreover, lithium plating on graphite anodes during low-temperature or high-rate charging can damage the SEI, promote secondary SEI formation, drive dendrite growth, degrade capacity, and increase resistance \cite{zhang2021lithiumplating}. These limitations have stimulated extensive research on alternative anode materials that can provide higher storage capacity, faster ion transport, and favorable low-voltage operation. In this context, two-dimensional (2D) materials have emerged as promising electrode candidates because of their large surface-to-mass ratio, abundant exposed active sites, short ion-diffusion pathways, and tunable electronic structures \cite{zhang2016beyond}. Recent studies have shown that 2D anode materials and their composites can offer improved ion accessibility and enhanced electrochemical performance compared with many conventional bulk materials \cite{xiao2021recent}. Several 2D materials, including graphene, borophene, phosphorene, MXenes, transition-metal carbides/nitrides, and transition metal dichalcogenides (TMDs), have therefore been explored as potential anodes for rechargeable metal-ion batteries. For example, borophene exhibits promising anode characteristics with Li and Na diffusion barriers of 0.6 and 0.3~eV, respectively, average open-circuit voltage ranges of 1.26--0.61~V for LIBs and 0.88--0.42~V for SIBs, and high theoretical capacities for Li and Na storage \cite{borophene}.  Researchers have also investigated phosphorene as a high-capacity anode material because of its favorable adsorption and anisotropic diffusion properties \cite{zhang2017phosphorene}. Similarly, MXenes and MBenes have attracted broad interest due to their metallic conductivity, surface tunability, and layered ion-storage channels \cite{naguib2014mxenes, ABI2025161270}.

Among these candidates, TMDs have attracted particular attention because
of their layered structure and rich chemical tunability. A TMD monolayer
(MX$_2$) consists of a transition-metal layer sandwiched between two
chalcogen layers, and the weak van der Waals interlayer interaction in
bulk TMDs enables exfoliation into monolayers \cite{pumera2014layered}.
The exposed monolayer surfaces provide accessible sites for alkali-metal
adsorption and migration \cite{lin2019tmd}, while the layered geometry
facilitates Li$^+$/Na$^+$ transport and allows the electronic and
electrochemical properties to be tuned at the atomic level
\cite{chen2020tmd}. Defect engineering, strain, heterostructure
formation, surface functionalization, and doping have been widely used
to tune the adsorption strength, conductivity, and diffusion kinetics of
TMD electrodes \cite{xu2017interlayer}. For instance, 1T-MoS$_2$ has
been proposed as an anode for Li-, Na-, and Mg-ion batteries
\cite{mos2}, and layered MoS$_2$ has shown a Na-induced 2H-to-1T phase
transformation, an electrode potential of 0.75--1.25~V, and a capacity
of 146~mAh~g$^{-1}$ as an SIB anode \cite{MORTAZAVI2014279}. Doped and
strained TiS$_2$ monolayers have shown improved alkali-metal adsorption
and diffusion \cite{nair2022tuning}, while WS$_2$/graphene
heterostructures have exhibited Li migration barriers of 0.24--0.28~eV
and a theoretical capacity of 588.16~mAh~g$^{-1}$
\cite{Bijoy2024WS2Graphene}. However, pristine TMDs often suffer from
semiconducting character, limited electronic conductivity, restacking,
and insufficient active-site utilization, making rational materials
engineering necessary \cite{gu2024molecular}.

Janus TMDs provide an additional degree of freedom for designing high-performance electrode materials. In a Janus monolayer, one chalcogen layer of a conventional TMD is replaced by a different chalcogen element, thereby breaking the out-of-plane mirror symmetry and generating two chemically distinct surfaces. The experimental realization of Janus MoSSe has demonstrated the feasibility of such asymmetric TMD monolayers \cite{lu2017janus}. This broken mirror symmetry can generate an intrinsic dipole moment across the out-of-plane direction \cite{tao2019dipole}. It can also modify the electronic, mechanical, magnetic, catalytic, and adsorption-related properties of the monolayer \cite{shi2018mechanical}. Such polarity is particularly attractive for battery electrodes because the two nonequivalent surfaces can interact differently with alkali-metal atoms, allowing surface-dependent adsorption and diffusion behavior. Theoretical studies have already predicted promising anode behavior for Janus TMDs. For example, Janus MoSSe was reported to exhibit favorable Li-ion storage with a theoretical capacity of 776.5~mAh~g$^{-1}$ and low migration barriers \cite{Shang2018MoSSe}. Janus WSSe has been reported to deliver theoretical capacities of 477.8, 371.5, and 156.0~mAh~g$^{-1}$ for Li, Na, and K storage, respectively, with corresponding average OCV values of 0.48, 0.57, and 0.37~V \cite{ahmad2023wsse}. These findings indicate that Janus asymmetry can be effectively exploited to tune adsorption strength, ion mobility, and electronic conductivity in 2D anode materials. Hafnium-based chalcogenides are also attractive within the TMD family because of their promising electronic and transport properties. In particular, few-layer HfS$_2$ has been experimentally demonstrated as an ultrasensitive phototransistor material with strong photoresponse \cite{xu2015hfs2}. Vertically oriented few-layer HfS$_2$ nanosheets have also been synthesized and characterized, further confirming the experimental accessibility of Hf-based layered chalcogenides \cite{zheng2016hfs2}. In addition, 2D HfS$_2$ has been theoretically reported as a promising photocatalyst for water-splitting applications \cite{singh2016hfs2}. More recently, first-principles studies have shown that Li and Na adsorption and diffusion on HfS$_2$ can be improved through nonmetal doping and strain engineering \cite{nayak2025hfs2}. These reports suggest that Hf-based TMDs can be tuned for energy-related applications. However, compared with other Janus TMDs, the electrochemical behavior of Janus HfSSe remains comparatively underexplored, especially for Li-ion and Na-ion battery anodes.

In this paper, pristine and Co-doped Janus HfSSe monolayers were systematically investigated using first-principles density functional theory (DFT) calculations to evaluate their potential as anode materials for Li-ion and Na-ion batteries. First, the pristine HfSSe monolayer was structurally optimized, and its dynamical stability was examined through phonon dispersion calculations. Subsequently, Co doping was introduced into the HfSSe supercell, and the structural and electronic properties of the doped system were analyzed. The adsorption behavior of Li and Na atoms on both pristine and Co-doped HfSSe was then investigated by considering different adsorption sites and layered adsorption configurations. Charge transfer, open-circuit voltage, theoretical specific capacity, and ion diffusion barriers were calculated to assess the electrochemical performance of the proposed anode materials. In addition, the electronic properties after the first Li and Na adsorption layers were examined to evaluate the effect of alkali-metal adsorption on the electronic characteristics of the monolayers. Finally, the calculated properties of pristine and Co-doped HfSSe were compared to assess the influence of Co doping on their suitability as anode materials for Li-ion and Na-ion batteries.

\begin{figure*}[width=\textwidth,pos=!t]
    \centering
    \includegraphics[width=1\textwidth]{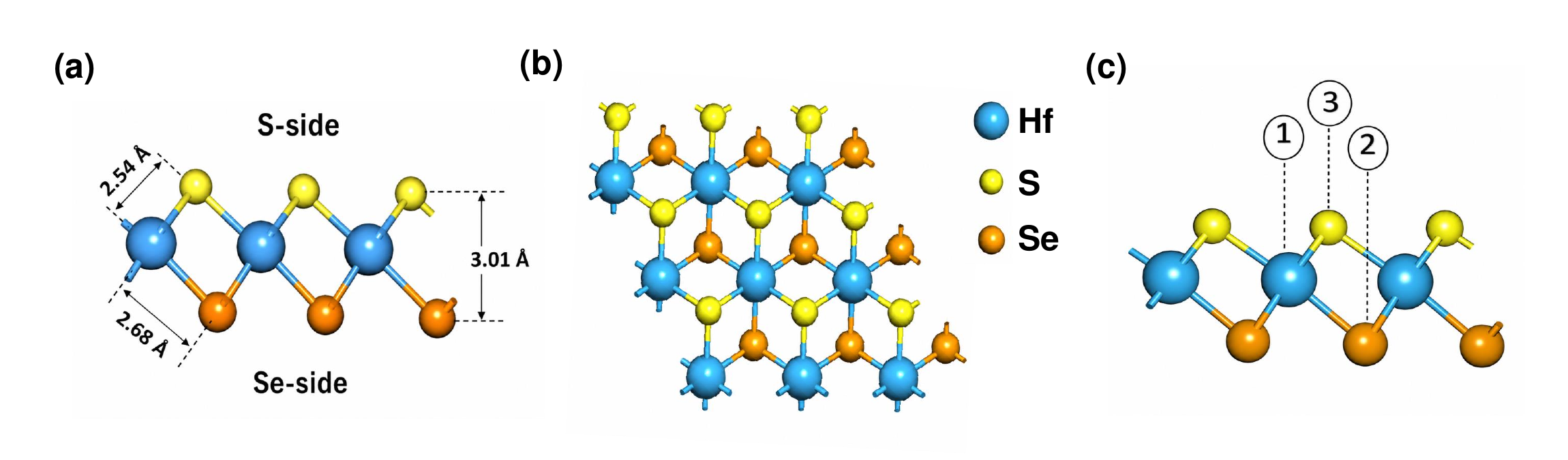}
    \caption{Optimized structure of the pristine HfSSe monolayer: (a) side view, (b) top view, and (c) possible adsorption sites considered for Li and Na adsorption.}
    \label{fig:structure_undoped}
\end{figure*}

\section{Computational details}

First-principles simulations were carried out using the \textsc{Quantum ESPRESSO} package \cite{Giannozzi_2009} within the framework of density functional theory (DFT). The interaction between valence electrons and ionic cores was described using projector-augmented-wave-type pseudopotentials, while the exchange-correlation energy was treated using the Perdew-Burke-Ernzerhof (PBE) functional within the generalized gradient approximation (GGA). Long-range dispersion interactions were included through the Grimme DFT-D3 correction \cite{grimme2010consistent}. A plane-wave kinetic-energy cutoff of 60 Ry and a charge-density cutoff of 600 Ry were used for all calculations. Gaussian smearing with a smearing width of 0.005 Ry was applied. The Brillouin zone was sampled using a Monkhorst-Pack ($16 \times 16 \times 1$) k-point mesh for structural optimization and self-consistent field calculations. The self-consistent electronic convergence threshold was set to $1.0 \times 10^{-10}$ Ry, while the total-energy and force convergence thresholds for relaxation were chosen as $1.0 \times 10^{-7}$ Ry and $1.0 \times 10^{-6}$ Ry/Bohr, respectively. The plane-wave kinetic-energy cutoff and $k$-point mesh were systematically optimized through convergence tests prior to the production calculations (see Supplementary Material for details). 

The optimized HfSSe unit cell was first obtained through a variable-cell relaxation procedure, in which both the atomic coordinates and lattice parameters were allowed to relax until the prescribed energy and force convergence criteria were satisfied. The relaxed unit cell was subsequently expanded to construct a ($3 \times 3 \times 1$) supercell, which was used for the subsequent adsorption and diffusion calculations. About 20~\AA\ of vacuum region in each side was maintained along the out-of-plane direction to eliminate spurious interactions between periodically repeated images of the monolayer. The dynamical stability of the relaxed monolayer was examined through phonon dispersion calculations using the finite-displacement method as implemented in the \textsc{PHONOPY} package ~\cite{Togo_2023}. A ($4 \times 4 \times 1$) supercell was used for the phonon calculations to obtain a reliable description of the vibrational modes of the two-dimensional structure. In the electronic property calculations, electronic band-structure calculations were performed along the high-symmetry path $\Gamma$--M--K--$\Gamma$ of the hexagonal Brillouin zone, sampled with 32 points between successive high-symmetry points. Density of states (DOS) and projected density of states (PDOS) calculations were carried out using a denser ($28 \times 28 \times 1$) k-point mesh in the non-self-consistent field calculation to obtain a smoother and more resolved electronic spectrum. The tetrahedron method was employed for the occupation scheme in the DOS calculations. To evaluate the ion migration behavior, diffusion pathways of the adsorbed atom were investigated using the climbing-image nudged elastic band method. In addition, Bader charge analysis was performed to quantify the charge transfer of the adsorbed systems \cite{bader}. Layer-by-layer adsorption calculations were performed by progressively increasing the Li/Na concentration until further adsorption became energetically unfavorable, thereby determining the maximum stable loading and the corresponding theoretical specific capacity. To evaluate the thermodynamic stability of the intermediate adsorption
configurations, formation-energy convex hulls were constructed. The stable
configurations lying on the lower convex hull were identified, and the
open-circuit voltage (OCV) was subsequently calculated from the total-energy
differences between successive stable adsorption states. The migration behavior of the adsorbed Li and Na atoms was investigated using the climbing-image nudged elastic band (CI-NEB) method.

\section{Results and discussion}

\subsection{Structure and stability}

\begin{figure*}[width=\textwidth,pos=!t]
    \centering
    \includegraphics[width=1\textwidth]{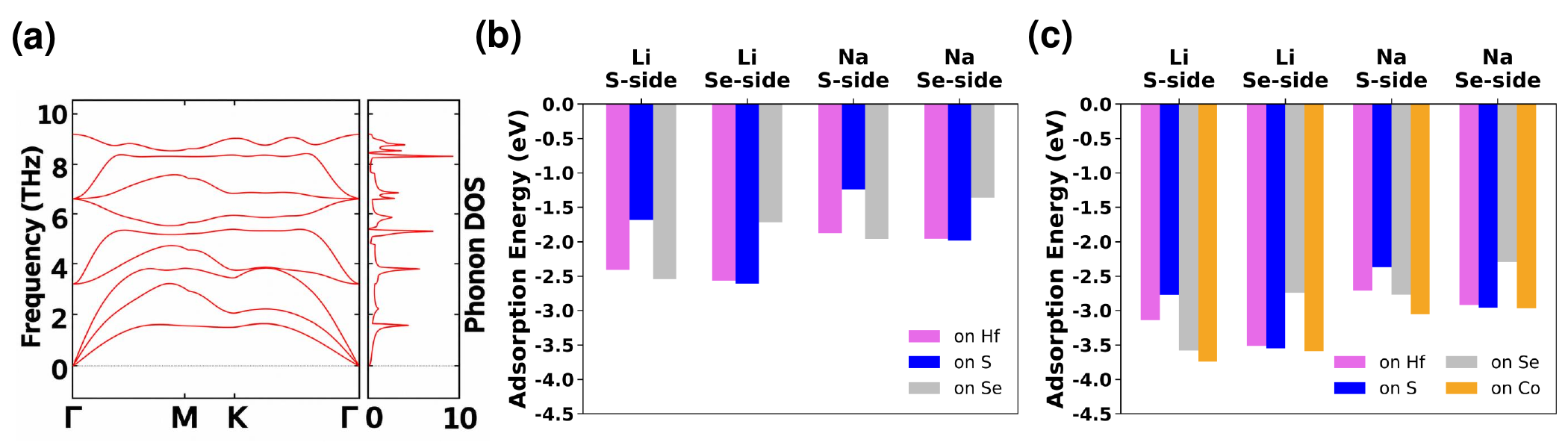}
 \caption{Phonon and adsorption characteristics of Janus HfSSe:
(a) phonon dispersion and phonon density of states of pristine HfSSe,
and adsorption energies of Li and Na atoms at different surface sites
of (b) pristine HfSSe and (c) Co-doped HfSSe.}
    \label{fig:chart}
\end{figure*}

The pristine HfSSe monolayer was constructed in a 1T Janus configuration with a hexagonal lattice. Due to the broken out-of-plane mirror symmetry caused by the presence of different chalcogen atoms on the two opposite surfaces, the optimized structure belongs to the trigonal space group $P3m1$. In this Janus structure, the Hf atomic layer is sandwiched between the S and Se atomic layers, forming an asymmetric S--Hf--Se arrangement. Accordingly, the two exposed surfaces of the monolayer are referred to as the S-side and Se-side, respectively. The optimized in-plane lattice constants were found to be $a=b=3.683$~\AA. The calculated Hf--S and Hf--Se bond lengths were approximately 2.54~\AA{} and 2.68~\AA, respectively. These values are consistent with the literature \cite{lattice3.68}. The vertical separation between the outer S and Se atomic planes, which represents the effective thickness of the Janus monolayer, is about 3.01~\AA. The side and top views of the optimized pristine HfSSe monolayer are shown in Fig.~\ref{fig:structure_undoped}(a) and Fig.~\ref{fig:structure_undoped}(b), respectively. The dynamical stability of the optimized pristine HfSSe monolayer was examined using phonon dispersion calculations. The phonon spectrum was calculated using the finite-displacement method as implemented in the PHONOPY package. As shown in Fig.~\ref{fig:chart}(a), no imaginary phonon frequency appeared along the selected high-symmetry path of the Brillouin zone. This confirmed the dynamical stability of the optimized monolayer structure. The corresponding phonon density of states also showed that all vibrational modes lay in the positive-frequency region, further supporting the stability of pristine HfSSe.

To evaluate the adsorption behavior of alkali metal atoms on pristine HfSSe, Li and Na atoms were placed at different possible adsorption sites on both the S-side and Se-side surfaces shown in Fig.~\ref{fig:structure_undoped}(c). For each adsorption configuration, the atomic positions were fully relaxed and the adsorption energy was calculated to determine the most favorable adsorption site. A negative value of adsorption energy indicates that the adsorption process is energetically favorable. The calculated adsorption energies for Li and Na atoms on different sites are shown in Fig.~\ref{fig:chart}(b). All adsorption energies are negative, indicating that both Li and Na atoms can be stably adsorbed on the pristine HfSSe monolayer.

\begin{figure*}[width=\textwidth,pos=!t,align=\centering]
    \centering
    \includegraphics[width=1\textwidth]{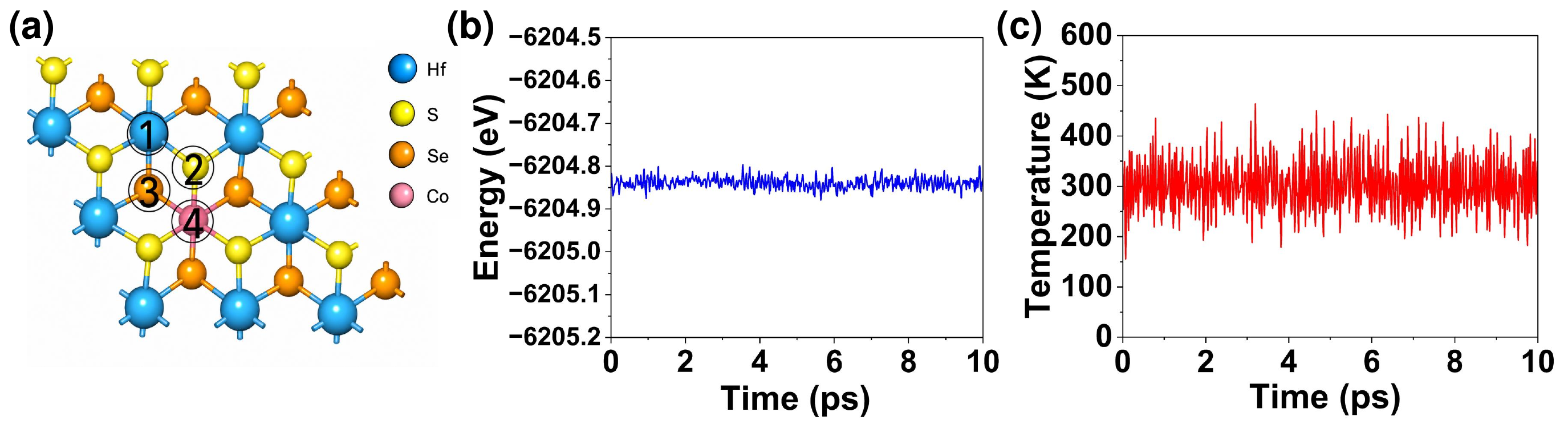}
    \caption{Co-doped HfSSe structure and AIMD results at 300~K for 10~ps: (a) optimized top view, (b) total energy variation, and (c) temperature fluctuation.}
    \label{fig:aimd_doped}
\end{figure*}

To further improve the adsorption performance of the HfSSe monolayer, Co doping was introduced by substituting one Hf atom with a Co atom in the ($3 \times 3 \times 1$) supercell. To identify an effective transition-metal dopant, the Li adsorption energies of Co-, Cu-, V-, Mn-, and Fe-doped HfSSe were compared. The calculated adsorption energies are $-3.7405$, $-3.4867$, $-2.8165$, $-3.3680$ and $-3.5490$~eV for the Co-, Cu-, V-, Mn-, and Fe-doped systems, respectively. The more negative value obtained for Co-doped HfSSe indicates stronger Li--surface interaction than in the other transition metal doped structures. Therefore, Co-doped HfSSe was selected for the detailed investigation. After the doping, the optimized structure is shown in Fig.~\ref{fig:aimd_doped}(a). The substitutional doping formation energy was calculated by,
\begin{equation}
E_f = E_{\mathrm{Co\mbox{-}HfSSe}} - E_{\mathrm{HfSSe}} + \mu_{\mathrm{Hf}} - \mu_{\mathrm{Co}},
\label{eq:formation_energy}
\end{equation}
where $E_{\mathrm{Co\mbox{-}HfSSe}}$ and $E_{\mathrm{HfSSe}}$ are the total energies of the Co-doped and pristine HfSSe monolayers, respectively. Here, $\mu_{\mathrm{Hf}}$ and $\mu_{\mathrm{Co}}$ represent the chemical potentials of Hf and Co, respectively. The calculated formation energy for Co substitution in place of Hf was found to be 3.44~eV. The calculated formation energy of Co-doped HfSSe is positive, indicating that the substitution of Co at the Hf site requires energy input under equilibrium conditions. However, a positive formation energy does not necessarily exclude experimental feasibility, since several transition-metal-doped TMD systems with positive defect or dopant formation energies have been experimentally realized. A similar situation was reported for transition-metal-doped TiS$_2$, where the calculated formation energies of Mn-, Fe-, Co-, Ni-, and Cu-doped TiS$_2$ were positive \cite{nair2022tuning}. However, experimental studies demonstrated the successful incorporation of transition-metal atoms such as Co \cite{zhang2011thermoelectric}, Ni \cite{Zhang_2006} into TiS$_2$-based layered compounds. Furthermore, transition-metal doping was widely demonstrated in other TMDs. For example, Co-doped MoS$_2$ nanosheets were experimentally synthesized and characterized, showing that cobalt incorporation can modify the structural, magnetic, optical, and catalytic properties of MoS$_2$ \cite{CoMos2}. Similarly, V-doped MoS$_2$ monolayers were prepared through chemical vapor deposition, where V atoms were confirmed to substitute Mo sites in the MoS$_2$ lattice \cite{zheng2023vapor}. Nb doping was also experimentally achieved in monolayer WS$_2$ \cite{nbws2} and WSe$_2$ \cite{nbwse2}, leading to effective p-type behavior and tunable electronic properties. These experimental reports suggested that transition-metal doping is a feasible strategy for modifying the properties of layered TMDs.

\begin{figure*}[t]
    \centering
    \includegraphics[width=1\textwidth]{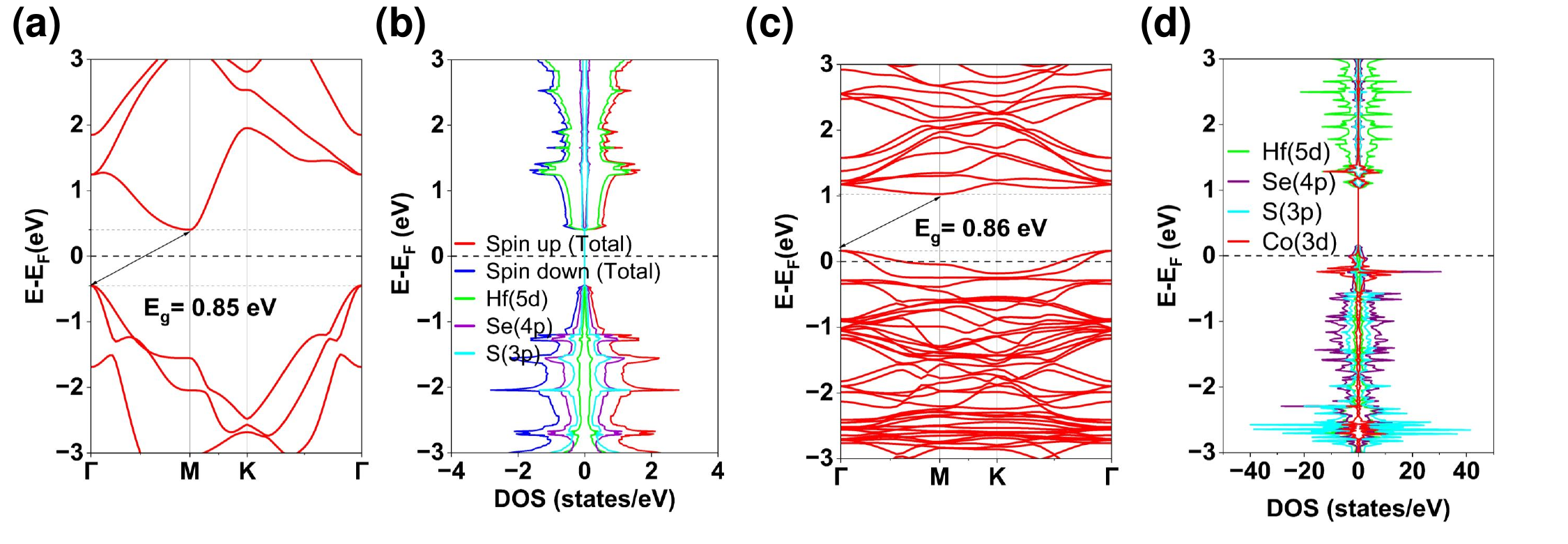}
    \caption{
(a) Band structure of pristine HfSSe (b) PDOS of pristine HfSSe (c)  band structure of Co-doped HfSSe (d) PDOS of Co-doped HfSSe.
}
    \label{fig:electronic_properties}
\end{figure*}

The adsorption behavior of Li and Na atoms on the Co-doped HfSSe monolayer was then investigated by placing the metal atoms at the selected adsorption sites on both the S-side and Se-side surfaces Fig.~\ref{fig:aimd_doped}(a). The calculated adsorption energies are presented in Fig.~\ref{fig:chart}(c). All adsorption energies are negative, indicating that Li and Na adsorption on the Co-doped HfSSe monolayer is energetically favorable. Compared with the pristine HfSSe monolayer, the adsorption energies become more negative after Co doping, which indicates that Co doping strengthens the binding of Li and Na atoms and can therefore improve the electrochemical performance of the HfSSe monolayer as an anode material.

The thermal stability of the Li-adsorbed Co-doped HfSSe monolayer was further examined using \textit{ab initio} molecular dynamics (AIMD) simulation. The simulation was performed for 10~ps at a target temperature of 300~K. As shown in Fig.~\ref{fig:aimd_doped}(b,c), the total energy fluctuated
within a narrow range without any continuous upward or downward drift
during the simulation. In addition, the temperature oscillated around
the target value of 300~K. These results indicated that the Li-adsorbed Co-doped HfSSe monolayer maintains structural stability under thermal conditions, supporting its suitability for battery anode applications.


\subsection{Electronic properties}

The electronic properties of pristine and Co-doped Janus HfSSe monolayers were investigated through electronic band-structure and projected density of states (PDOS) calculations, as shown in Fig.~\ref{fig:electronic_properties}. The energy zero in all plots is referenced to the Fermi level, $E_{\mathrm{F}}$. The pristine HfSSe monolayer exhibited semiconducting behavior with an indirect band gap of approximately 0.85~eV, which was in agreement
with the previous report \cite{BERA2021159704}. The valence-band maximum (VBM) was located at the $\Gamma$ point, whereas the conduction-band minimum (CBM) occurred near the M point along the $\Gamma$--M--K--$\Gamma$ high-symmetry path. Although this finite band gap was favorable for electronic applications, the absence of charge carriers at $E_{\mathrm{F}}$ could limit the intrinsic electronic conductivity of pristine HfSSe during battery operation.

\begin{figure*}[t]
    \centering
    \includegraphics[width=\textwidth]{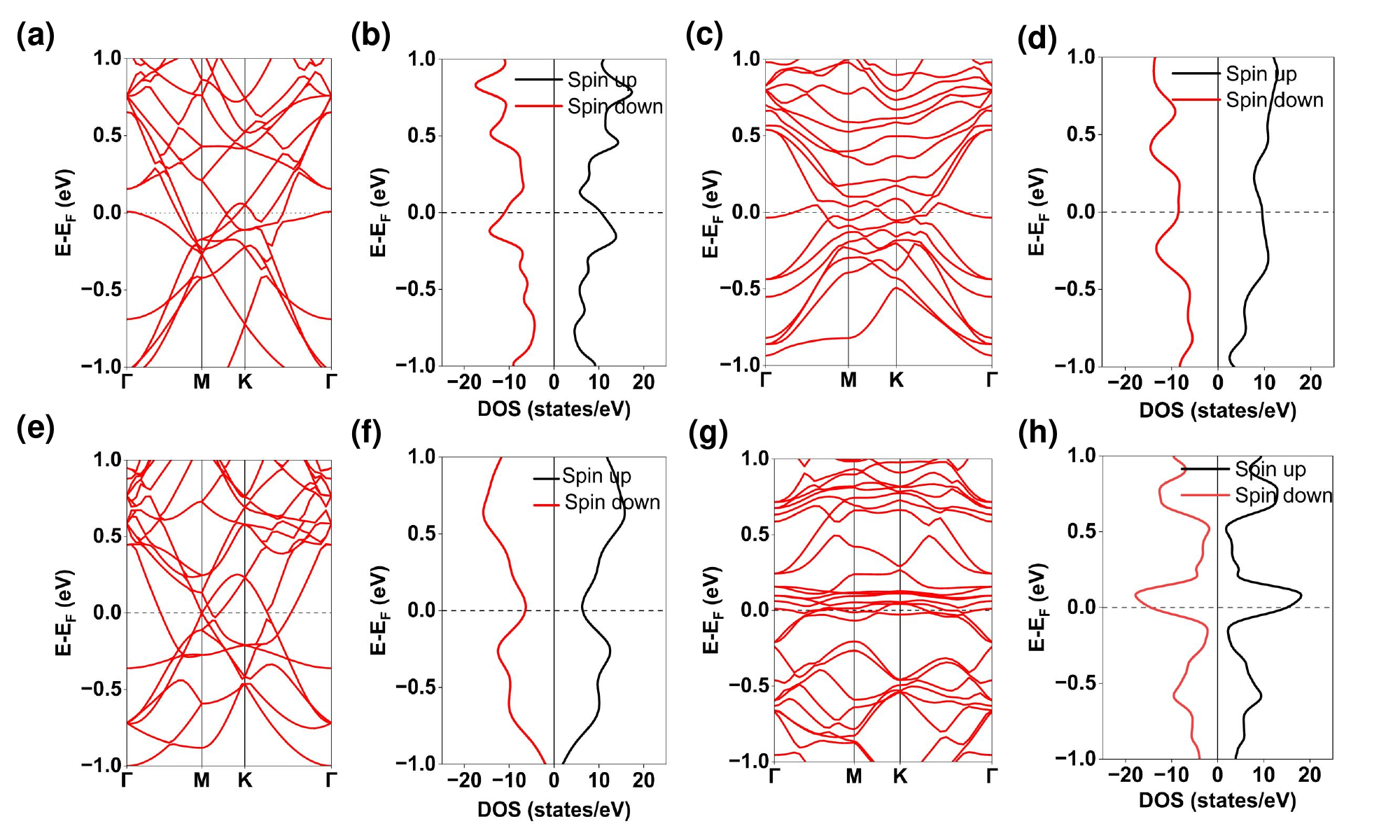}
    \caption{Spin-polarized band structures and total density of states after
    first-layer adsorption: (a,b) Li on pristine HfSSe, (c,d) Li on
    Co-doped HfSSe, (e,f) Na on pristine HfSSe, and (g,h) Na on
    Co-doped HfSSe. The Fermi level is set to 0~eV.}
    \label{fig:first_layer_band_dos}
\end{figure*}
The PDOS analysis further clarified the orbital contributions to the
electronic states near the band edges. The upper valence-band region was
predominantly composed of Se-$4p$ and S-$3p$ states, with the Se-$4p$
contribution being more pronounced near the valence-band maximum.
In contrast, the conduction-band edge was mainly dominated by Hf-$5d$
states, with additional hybridization from the Se-$4p$ and S-$3p$
orbitals. Therefore, the electronic states across the band gap were
primarily characterized by chalcogen-derived Se-$4p$/S-$3p$ states in
the valence band and Hf-$5d$ states in the conduction band.

The electronic properties of the Co-doped HfSSe monolayer were also
investigated. As shown in Figs.~\ref{fig:electronic_properties}(c,d), the
Co-doped system preserved a finite separation of approximately 0.86~eV
between the valence- and conduction-band edges, which was essentially
unchanged from the 0.85~eV value obtained for pristine HfSSe. Hence, Co
incorporation did not significantly alter the fundamental band-edge
separation. The major electronic effect of Co substitution was a downward
shift of the Fermi level into the upper valence-band region. This behavior
indicated acceptor-like charge redistribution in the Co-substituted system, resulting in partially unoccupied states, or holes, within the
chalcogen-$p$-dominated valence manifold. Therefore, the Co-doped
monolayer was more appropriately described as a degenerate $p$-type
semiconductor rather than an intrinsic semiconductor. The presence of
accessible valence-band carriers was expected to improve electronic
transport relative to pristine HfSSe, which is beneficial for anode
conductivity. The PDOS shown in Fig.~\ref{fig:electronic_properties}(d) indicates that Co-$3d$ states contributed significantly to the electronic states near the valence-band edge and also participated in states close to the conduction-band region. The Co-doped HfSSe monolayer exhibited weak
magnetic character, as evidenced by the slight asymmetry between the
spin-up and spin-down electronic states, indicating spin polarization
induced by Co substitution. Replacing Hf with Co produced a local
S$_3$CoSe$_3$ coordination environment, which modified the crystal-field
environment around the dopant. Consequently, the Co-$3d$ orbitals
underwent crystal-field splitting and hybridized with the neighboring
S-$3p$, Se-$4p$, and Hf-$5d$ states. This Co-$3d$/chalcogen-$p$
hybridization modified the electronic states near the band edges and was
consistent with the observed shift of the Fermi level toward the
valence-band region. The Co-derived states also increased the availability
of electronic states near the Fermi level, which could facilitate charge
transport during Li and Na adsorption. It should be noted that the larger number of bands visible in Fig.~\ref{fig:electronic_properties}(c) partly results from Brillouin-zone folding associated with the larger ($3\times3\times1$) supercell. Therefore, the electronic effect of Co substitution was interpreted primarily from the Fermi-level shift and the Co-$3d$-related PDOS features rather than from the apparent increase in the number of bands. Overall, Co substitution preserved the basic valence--conduction band separation of Janus HfSSe while introducing Co-$3d$-derived states, weak spin polarization, and a higher availability of electronic states near the Fermi level. To further examine the contribution of Li to the electronic structure, PDOS calculations were also performed after adsorption of a single Li atom on the HfSSe monolayers (see Supplementary Material).

The spin-polarized band structures and TDOS after first-layer Li and Na
adsorption are presented in Fig.~\ref{fig:first_layer_band_dos}. The pristine adsorbed structures exhibited nearly
symmetric spin-up and spin-down states, indicating negligible magnetism.
In contrast, the Co-doped configurations showed a visible asymmetry
between the two spin channels, indicating that weak spin polarization
was retained after Li and Na adsorption due to the Co-derived electronic
states. For all four configurations, several bands crossed the Fermi
level and the TDOS remained finite at $E_{\mathrm{F}}$. Therefore,
adsorption of the first Li or Na layer converted both pristine and
Co-doped HfSSe into metallic systems. This metallization was associated
with charge transfer from the alkali-metal atoms to the host monolayer,
which introduced partially filled states near the Fermi level. The
resulting increase in the availability of charge carriers was expected
to improve electronic conductivity, facilitate electron transport during
lithiation and sodiation, and thereby support improved rate performance
of the anode material.

\subsection{Adsorption properties and charge density}

To assess the suitability of pristine and Co-doped Janus HfSSe monolayers as potential anode materials, the adsorption behavior of single Li and Na atoms was first examined at different possible surface sites. Since HfSSe is a Janus monolayer with two chemically nonequivalent surfaces, adsorption was examined separately on the S-side and Se-side surfaces. For the pristine monolayer, three representative adsorption sites were considered: site 1 on top of Hf, site 2 on top of S, and site 3 on top of Se shown in Fig.~\ref{fig:structure_undoped}(c). For the Co-doped monolayer, an additional site was considered directly above the substituted Co atom, as shown in Fig.~\ref{fig:aimd_doped}(a). For each configuration, the atomic positions were fully relaxed before calculating the adsorption energy. The adsorption energy was calculated using,
\begin{equation}
E_{\mathrm{ads}} = E_{\mathrm{M+host}} - E_{\mathrm{host}} - E_{\mathrm{M}},
\label{eq:adsorption_energy}
\end{equation}
where $E_{\mathrm{M+host}}$ is the total energy of the Li- or Na-adsorbed system, $E_{\mathrm{host}}$ is the total energy of the pristine or Co-doped HfSSe monolayer, and $E_{\mathrm{M}}$ is the energy of an isolated Li or Na atom. A negative adsorption energy indicates that the adsorption process is energetically favorable, while a more negative value represents stronger interaction between the adsorbed atom and the monolayer surface.

\begin{table}[t]
\centering
\caption{Calculated adsorption energies of Li and Na atoms on pristine HfSSe monolayer}
\label{tab:ads_pristine}
\small
\setlength{\tabcolsep}{7pt}
\renewcommand{\arraystretch}{1.08}
\begin{tabular}{cccc}
\toprule
\textbf{Ion} & \textbf{Site} & \textbf{S side (eV)} & \textbf{Se side (eV)}\\
\midrule
\multirow{3}{*}{Li} 
& on Hf & -2.4085 & -2.5652 \\
& on S & -1.6850 & -2.6125 \\
& on Se & -2.5452 & -1.7216 \\
\midrule
\multirow{3}{*}{Na} 
& on Hf & -1.8769 & -1.9578 \\
& on S & -1.2413 & -1.9800 \\
& on Se & -1.9609 & -1.3621 \\
\bottomrule
\end{tabular}
\end{table}
\begin{table}[t]
\centering
\caption{Calculated adsorption energies of Li and Na atoms on Co-doped HfSSe monolayer}
\label{tab:ads_codoped}
\small
\setlength{\tabcolsep}{7pt}
\renewcommand{\arraystretch}{1.08}
\begin{tabular}{cccc}
\toprule
\textbf{Ion} & \textbf{Site} & \textbf{S side (eV)} & \textbf{Se side (eV)} \\
\midrule
\multirow{4}{*}{Li} 
& on Hf & -3.1404 & -3.5138 \\
& on S & -2.7729 & -3.5492 \\
& on Se & -3.5791 & -2.7464 \\
& on Co & -3.7405 & -3.5874 \\
\midrule
\multirow{4}{*}{Na} 
& on Hf & -2.7145 & -2.9202 \\
& on S & -2.3693 & -2.9578 \\
& on Se & -2.7711 & -2.2939 \\
& on Co & -3.0529 & -2.9702 \\
\bottomrule
\end{tabular}
\end{table}

The calculated adsorption energies are summarized in
Table~\ref{tab:ads_pristine} for pristine HfSSe. All adsorption energies
were negative, confirming that both Li and Na atoms could be stably
adsorbed on the monolayer surface. For Li adsorption, the most favorable
configuration was obtained on the Se side at the S site, with an
adsorption energy of $-2.6125$~eV. On the S side, Li adsorption was also
strong at the Se site, with an adsorption energy of $-2.5452$~eV. For
Na adsorption, the most stable configuration was likewise obtained on
the Se side at the S site, with an adsorption energy of $-1.9800$~eV,
while the S-side Se site yielded a comparable adsorption energy of
$-1.9609$~eV. These results indicated that the preferred adsorption site
depended on the surface termination, which was consistent with the
chemically distinct local environments provided by the S- and
Se-terminated surfaces of the Janus monolayer. Compared with Na, Li
exhibited more negative adsorption energies on pristine HfSSe,
indicating stronger Li--surface interactions. This difference could be
associated with the smaller atomic size of Li and its stronger
interaction with the surface electronic charge distribution.

After Co doping, the adsorption strength increased significantly for
both Li and Na, as summarized in Table~\ref{tab:ads_codoped}. For Li
adsorption, the strongest interaction occurred at the Co site on the
S side, with an adsorption energy of $-3.7405$~eV. The Co site was also
the most favorable Li adsorption position on the Se side, with an
adsorption energy of $-3.5874$~eV. Similarly, Na adsorption was most
favorable at the Co site, with adsorption energies of $-3.0529$~eV and
$-2.9702$~eV on the S and Se sides, respectively. Therefore, the
substituted Co atom acted as a highly active adsorption center for
alkali-metal atoms. The enhancement in adsorption strength after Co
substitution could be attributed to the modified local electronic
environment around the dopant site. Replacing Hf with Co introduced
localized Co-$3d$ states and local structural distortion, which enhanced
the interaction between the adsorbed Li/Na atoms and the neighboring S
and Se atoms. As a result, Co doping strengthened alkali-metal binding
compared with pristine HfSSe. The strongest Li adsorption energy changed
from $-2.6125$~eV in pristine HfSSe to $-3.7405$~eV in Co-doped HfSSe,
while the strongest Na adsorption energy changed from $-1.9800$~eV to
$-3.0529$~eV. These changes represented a substantial enhancement in
adsorption strength and demonstrated that Co substitution effectively
improved the surface activity of HfSSe toward Li and Na adsorption.
\begin{figure*}[t]
    \centering
    \includegraphics[width=\textwidth]{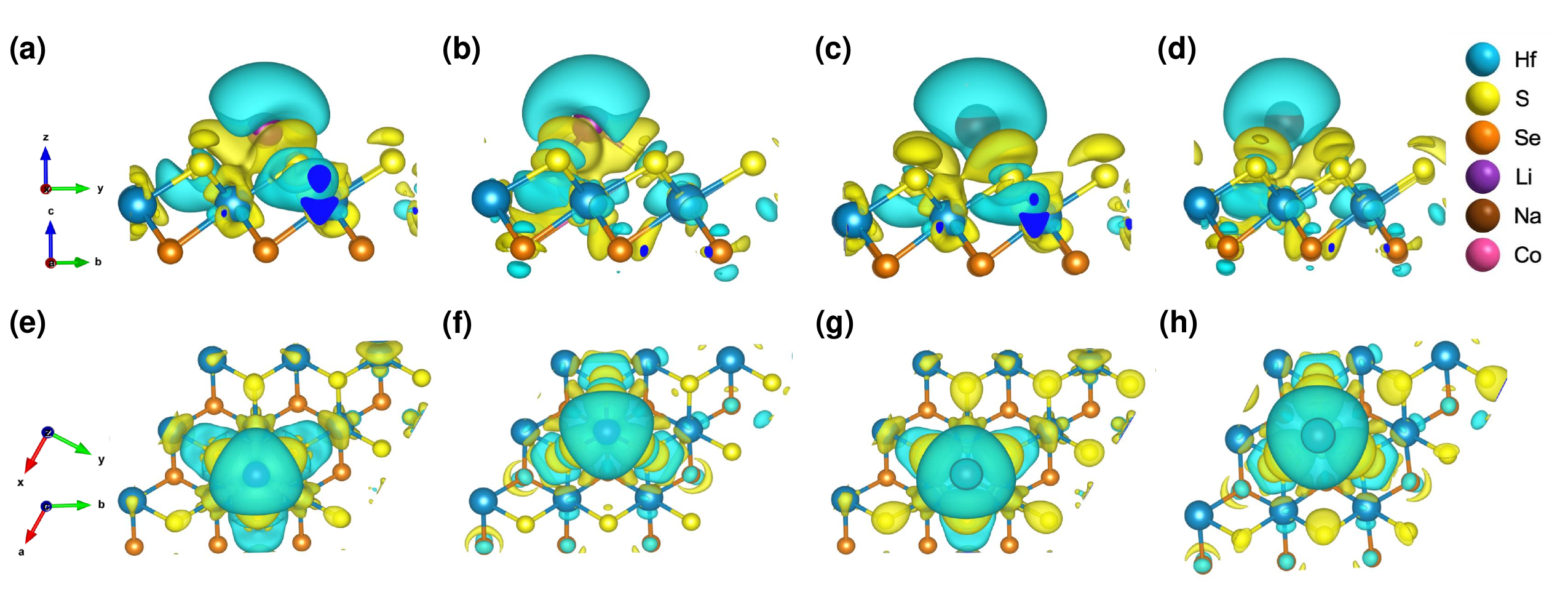}
    \caption{Charge density difference plots for Li and Na adsorption on pristine and Co-doped HfSSe monolayers. 
    (a,e) Li adsorption on pristine HfSSe, (b,f) Li adsorption on Co-doped HfSSe, 
    (c,g) Na adsorption on pristine HfSSe, and (d,h) Na adsorption on Co-doped HfSSe. 
    The cyan and yellow regions represent electron depletion and electron accumulation, respectively.}
    \label{fig:charge_density}
\end{figure*}

\begin{table*}[t]
\centering
\caption{Bader charges of Li- and Na-adsorbed pristine and Co-doped HfSSe
monolayers, grouped by local atomic environment. Positive (negative)
values denote electron loss (gain), and values are group averages. The
last column gives the number of atoms in each group for the
pristine/Co-doped cells.}
\label{tab:bader}
\small
\begin{tabular}{lccccc}
\toprule
 & \multicolumn{2}{c}{\textbf{Li adsorption}} & \multicolumn{2}{c}{\textbf{Na adsorption}} & \\
\cmidrule(lr){2-3}\cmidrule(lr){4-5}
\textbf{Atom group} & \textbf{Pristine} & \textbf{Co-doped} & \textbf{Pristine} & \textbf{Co-doped} & \textbf{Atoms} \\
 & $\boldsymbol{\Delta q}$ \textbf{(\textit{e})} & $\boldsymbol{\Delta q}$ \textbf{(\textit{e})} & $\boldsymbol{\Delta q}$ \textbf{(\textit{e})} & $\boldsymbol{\Delta q}$ \textbf{(\textit{e})} & \\
\midrule
M (Li/Na)             & $+0.8899$ & $+0.8883$ & $+0.8867$ & $+0.8769$ & 1/1 \\
Co                    & --        & $+0.5749$ & --        & $+0.5887$ & --/1 \\
Hf                    & $+2.0831$ & $+2.1168$ & $+2.0841$ & $+2.1171$ & 9/8 \\
S bonded to Co        & --        & $-1.1195$ & --        & $-1.0977$ & --/3 \\
S accepting M charge  & $-1.2794$ & --        & $-1.2597$ & --        & 3/-- \\
Other S               & $-1.1492$ & $-1.1524$ & $-1.1629$ & $-1.1654$ & 6/6 \\
Se bonded to Co       & --        & $-0.7372$ & --        & $-0.7391$ & --/3 \\
Other Se              & $-0.9894$ & $-0.9856$ & $-0.9874$ & $-0.9833$ & 9/6 \\

\bottomrule
\end{tabular}
\end{table*}

To further understand the interaction mechanism between the adsorbed
alkali-metal atoms and the HfSSe monolayer, charge density difference
(CDD) analysis was performed for Li and Na adsorption on both pristine
and Co-doped HfSSe. The charge density difference was calculated using,
\begin{equation}
\Delta \rho = \rho_{\mathrm{M+host}}-\rho_{\mathrm{host}}-\rho_{\mathrm{M}},
\end{equation}
where $\rho_{\mathrm{M+host}}$, $\rho_{\mathrm{host}}$, and
$\rho_{\mathrm{M}}$ represent the charge densities of the adsorbed
system, the pristine or Co-doped HfSSe monolayer, and the isolated Li/Na
atom, respectively. The resulting maps are presented in
Fig.~\ref{fig:charge_density}, where the cyan and yellow regions denote
electron depletion and accumulation, respectively. For both Li
Figs.~\ref{fig:charge_density}(a,e) and Na
Figs.~\ref{fig:charge_density}(c,g) on pristine HfSSe, electron
depletion was localized around the adsorbed atom, while electron
accumulation appeared between the adsorbate and the neighboring surface
atoms, indicating electron transfer from the alkali metal to the host.
After Co doping, the overall charge-transfer pattern was preserved, but
the charge redistribution became concentrated at the Co substitution
site and its neighboring chalcogen atoms, where lobed depletion and
accumulation features appeared Figs.~\ref{fig:charge_density}(b,d,f,h).
This indicated that the Co site and its surrounding chalcogen atoms
participated directly in accommodating the electron donated by the
adsorbate.
\begin{figure*}[t]
    \centering
    \includegraphics[width=1\textwidth]{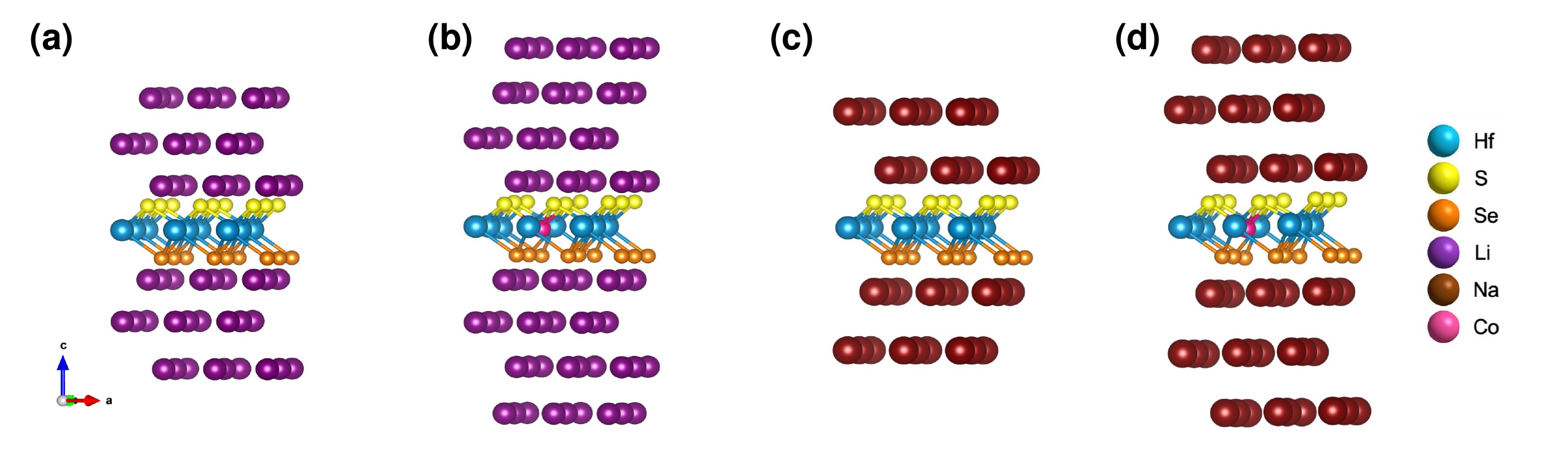}
    \caption{Optimized structures at the maximum stable alkali-metal loading:
    (a) Li-adsorbed pristine HfSSe, (b) Li-adsorbed Co-doped HfSSe,
    (c) Na-adsorbed pristine HfSSe, and (d) Na-adsorbed Co-doped HfSSe.}
    \label{fig:max_adsorption}
\end{figure*}
\begin{figure*}[t]
    \centering
    \includegraphics[width=0.98\textwidth]{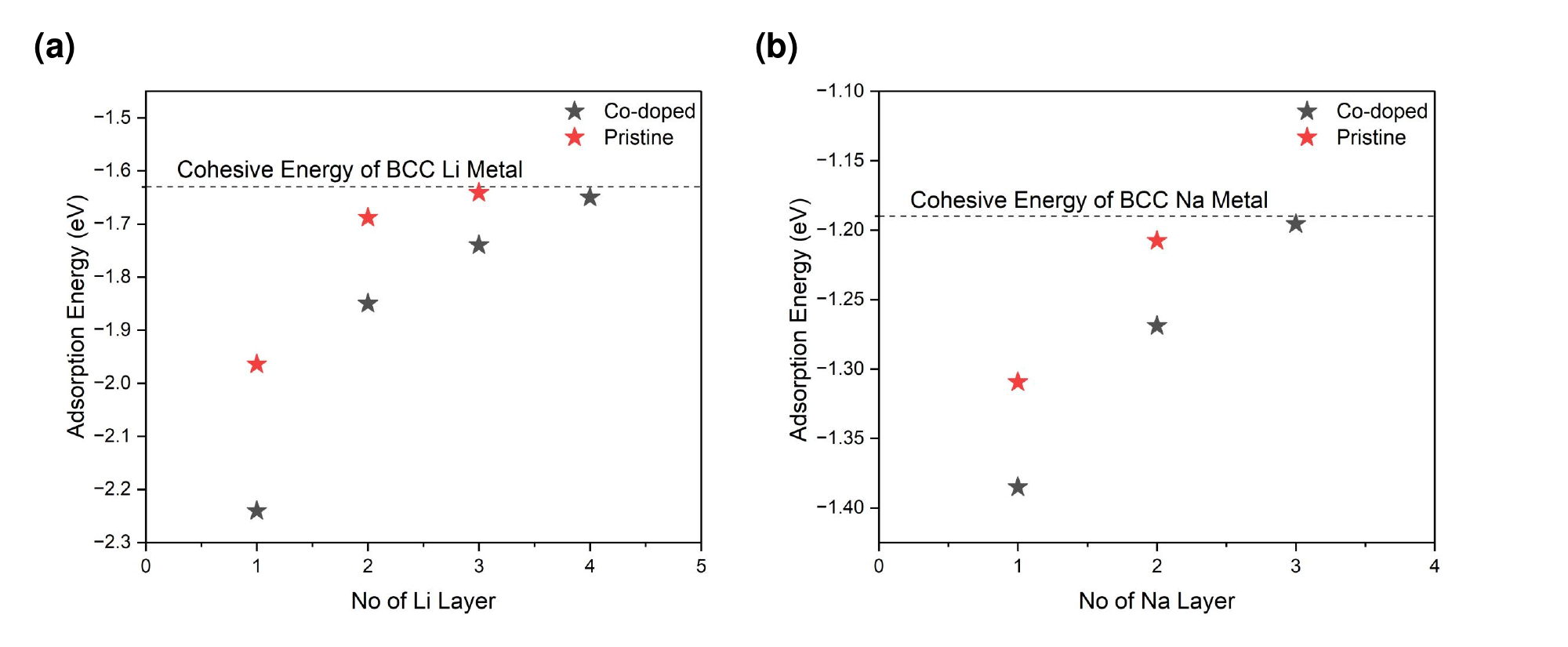}
    \caption{
    Layer-by-layer adsorption energies of (a) Li and (b) Na atoms on pristine and Co-doped HfSSe monolayers.
    }
    \label{fig:layered_adsorption_energy}
\end{figure*}

To quantify the charge transfer, Bader charge analysis was performed,
and the results are summarized in Table~\ref{tab:bader}. The adsorbed Li
and Na atoms lost 0.88--0.89~$e$ on both pristine and Co-doped HfSSe.
Since each atom has a single valence electron, these values showed that
Li and Na were almost completely ionized and existed on the surface
essentially as Li$^{+}$ and Na$^{+}$, with the donated electron stored
in the host monolayer. This corresponds to the anode operation, in which
the alkali-metal ion is supplied by the electrolyte and its compensating
electron is accommodated by the host through the external circuit; the adsorbed system allows these
electrons to be conducted efficiently.

Notably, the charge transfer from Li and Na remained nearly unchanged or even slightly decreased after Co doping, so the stronger adsorption
on Co-doped HfSSe could not be attributed to a larger charge transfer.
Instead, it originated from the energy of the states that accepted the
donated electron. Each Hf atom donated about 2.1~$e$ to its chalcogen
neighbors, whereas the substituted Co atom donated only about 0.58~$e$,
consistent with the higher electronegativity of Co than Hf.
Consequently, the three Se atoms bonded to Co carried about $-0.74~e$,
compared with about $-0.98~e$ for Se atoms away from the dopant. This
local electron deficiency produced unoccupied chalcogen-$p$ states at
the top of the valence band, consistent with the shift of the Fermi
level into the valence band. In pristine HfSSe, the
filled valence band forced the donated electron into the higher-lying
Hf-$5d$ conduction-band states, whereas in Co-doped HfSSe it could be
accommodated in these lower-energy empty states. This is reflected in
the S atoms bonded to Co, which lie directly beneath the adsorbate and
showed only a small deficit relative to the other S atoms, indicating
that the donated electron largely refilled the depleted states around
the dopant. In addition, the smaller positive charge on Co reduced the
electrostatic repulsion between the cationic adsorbate and the metal
site beneath it. The similar enhancement for Li ($\sim$1.13~eV) and Na ($\sim$1.07~eV)
indicated that the strengthened binding was governed mainly by the
modified electronic states of the host rather than by the specific
adsorbate, and the localization of the charge redistribution around the
Co site in Fig.~\ref{fig:charge_density} was consistent with this
mechanism. Overall, the
combined CDD and Bader analyses confirmed the predominantly ionic nature
of Li/Na adsorption and identified the Co site as the most active
adsorption center, which could contribute to stable ion storage and
suppress metal clustering during battery operation.

\subsection{Theoretical specific capacity and open-circuit voltage}

High ion-storage capacity and suitable low operating voltage are essential
requirements for high-performance anode materials. Therefore, after identifying
the most favorable single-atom adsorption sites, layer-by-layer adsorption of
Li and Na atoms was investigated on both pristine and Co-doped HfSSe
monolayers. The optimized structures at the maximum stable Li and Na loadings are shown in Fig.~\ref{fig:max_adsorption}. The metal atoms were gradually adsorbed on both sides of the
monolayer until the average adsorption energy became less favorable than the
cohesive energy of the corresponding bulk metal. This criterion was used to
determine whether surface adsorption was energetically preferred over metal
clustering. The average adsorption energy for each adsorbed layer was
calculated as \cite{ahmad2023wsse},

\begin{equation}
E_{\mathrm{ads}}^{n}
=
\frac{
E_{\mathrm{host}+nM}
-
E_{\mathrm{host}}
-
nE_M
}
{n},
\label{eq:avg_adsorption}
\end{equation}

where $E_{\mathrm{host}+nM}$ represents the total energy of the HfSSe or
Co-doped HfSSe monolayer after adsorption of $n$ Li/Na atoms,
$E_{\mathrm{host}}$ represents the total energy of the bare monolayer, and
$E_M$ represents the energy per isolated atom of Li or Na. Here, $n$ denotes the
number of adsorbed metal atoms at a particular adsorption stage. Fig.~\ref{fig:layered_adsorption_energy} shows that Li adsorption remained
energetically favorable up to the third layer on pristine HfSSe and up to the fourth layer on Co-doped HfSSe. Therefore, pristine HfSSe accommodated a
maximum of 54 Li atoms, whereas Co-doped HfSSe accommodated up to 72 Li atoms. For Na adsorption, the pristine monolayer remained stable up to two layers, corresponding to 36 Na atoms, while the Co-doped system supported three stable Na layers, corresponding to 54 Na atoms. These results indicated that Co substitution improved the multilayer adsorption capability of HfSSe for both Li and Na storage.

\begin{figure*}[t]
    \centering
    \includegraphics[width=1\textwidth]{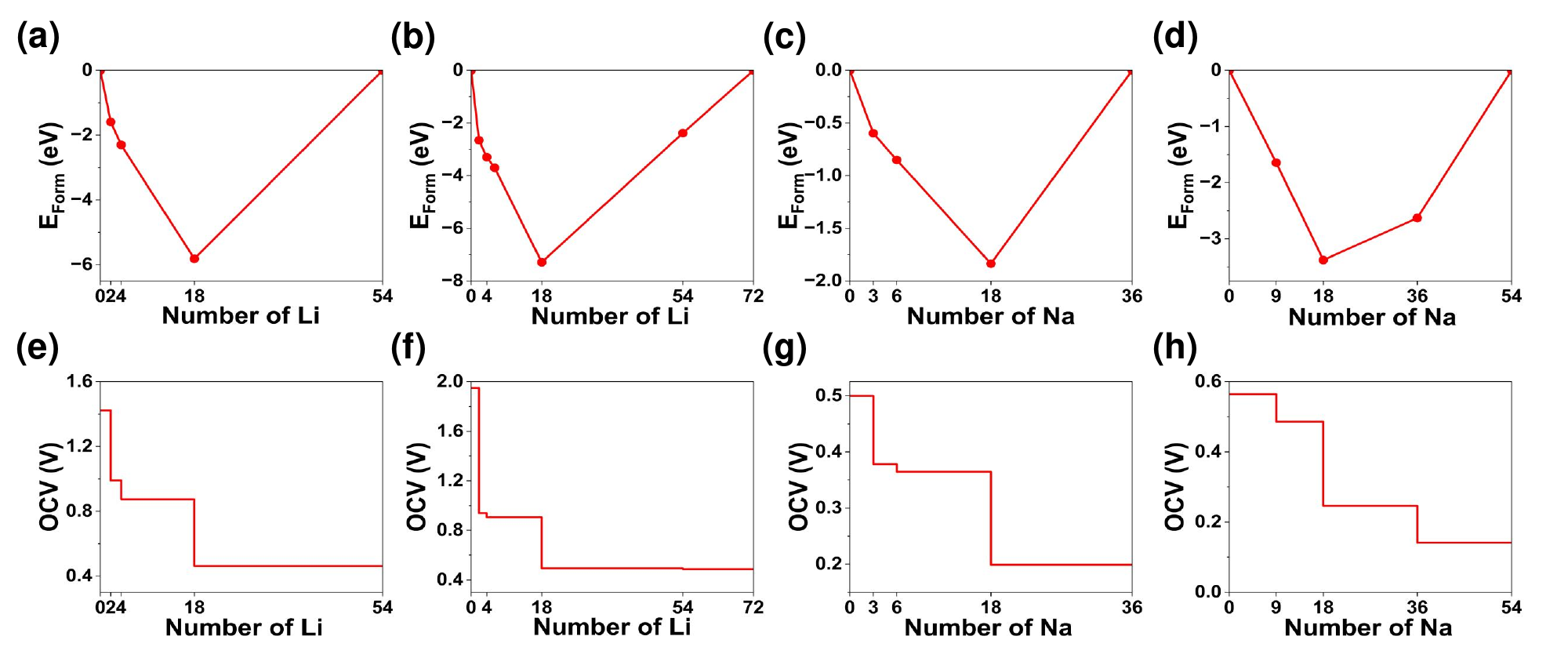}
    \caption{
    Formation-energy convex hulls and open-circuit voltage profiles of Li and Na adsorption on HfSSe monolayers: (a) Li formation energy of pristine HfSSe, (b) Li formation energy of Co-doped HfSSe, (c) Na formation energy of pristine HfSSe, (d) Na formation energy of Co-doped HfSSe, (e) Li OCV profile of pristine HfSSe, (f) Li OCV profile of Co-doped HfSSe, (g) Na OCV profile of pristine HfSSe, and (h) Na OCV profile of Co-doped HfSSe.
    }
    \label{fig:formation_ocv}
\end{figure*}
\begin{table}[t]
\centering
\caption{Calculated theoretical specific capacities of pristine and Co-doped HfSSe monolayers}
\label{tab:specific_capacity}
\small
\setlength{\tabcolsep}{5pt}
\renewcommand{\arraystretch}{1.08}
\begin{tabular}{llcc}
\toprule
\textbf{Ion} & \textbf{System} & \textbf{$N_{\mathrm{max}}$} & \textbf{Capacity (mAh g$^{-1}$)} \\
\midrule
\multirow{2}{*}{Li}
& Pristine HfSSe & 54 & 555.43 \\
& Co-doped HfSSe & 72 & 776.19 \\
\midrule
\multirow{2}{*}{Na}
& Pristine HfSSe & 36 & 370.29 \\
& Co-doped HfSSe & 54 & 582.14 \\
\bottomrule
\end{tabular}
\end{table}

The theoretical specific capacity was calculated from the maximum number of adsorbed metal atoms using Faraday's relation \cite{ABI2025161270},
\begin{equation}
C =
\frac{A N_{\mathrm{max}} F}{3.6M_{\mathrm{host}}},
\label{eq:specific_capacity}
\end{equation}
where $C$ is the theoretical specific capacity in mAh g$^{-1}$, $A$ is the number of electrons transferred by each adsorbed atom, $N_{\mathrm{max}}$ is the maximum number of adsorbed Li or Na atoms, $F$ is Faraday's constant, and $M_{\mathrm{host}}$ is the molar mass of the host structure. Since each Li or Na atom contributes one electron to the
electrochemical reaction, $A=1$ was used in the capacity calculation. The calculated capacities are summarized in Table~\ref{tab:specific_capacity}. For Li storage, the theoretical specific capacity increased from 555.43~mAh~g$^{-1}$ for pristine HfSSe to 776.19~mAh~g$^{-1}$ for Co-doped HfSSe. Similarly, for Na storage, the capacity increased from 370.29~mAh~g$^{-1}$ to 582.14~mAh~g$^{-1}$ after Co doping. The improvement originated from the larger number of Li and Na atoms that could be stably accommodated on the Co-doped monolayer.

To evaluate the electrochemical stability of the intermediate adsorption states, formation-energy convex hulls were constructed for Li and Na adsorption. The formation energy of each adsorbed configuration was calculated as \cite{formation},
\begin{equation}
E_{\mathrm{form}}(n)
=
E_{\mathrm{host}+nM}
-
\frac{n}{N_{\mathrm{max}}}E_{\mathrm{host}+N_{\mathrm{max}}M}
-
\frac{N_{\mathrm{max}}-n}{N_{\mathrm{max}}}E_{\mathrm{host}},
\label{eq:formation_energy_ads}
\end{equation}
where $E_{\mathrm{host}+nM}$ is the total energy of the host after adsorption of $n$ metal atoms, $E_{\mathrm{host}+N_{\mathrm{max}}M}$ is the total energy of the fully adsorbed structure, and $E_{\mathrm{host}}$ is the total energy of the bare pristine or Co-doped HfSSe monolayer. The points located on the lower convex hull were considered thermodynamically stable and were used for the OCV calculation. The open-circuit voltage between two stable adsorption states containing $p$ and $y$ metal atoms was calculated using \cite{ocv},
\begin{equation}
V =
-
\frac{
E_{\mathrm{host}+yM}
-
E_{\mathrm{host}+pM}
-
(y-p)E_M
}
{(y-p)e},
\label{eq:ocv}
\end{equation}
where $E_{\mathrm{host}+yM}$ and $E_{\mathrm{host}+pM}$ are the total energies of the adsorbed structures containing $y$ and $p$ metal atoms, respectively. Here, $E_M$ is the energy of one Li or Na atom in the bulk metal phase, and $e$ is the elementary charge. Since the total energies are expressed in eV, the voltage is obtained directly from the energy difference per transferred electron.

As shown in Fig.~\ref{fig:formation_ocv}, the voltage decreases stepwise
with increasing Li or Na concentration, which is typical of layer-by-layer
adsorption on two-dimensional anode materials. For Li adsorption, both
pristine and Co-doped HfSSe exhibited an average OCV of approximately
0.62~V. For Na adsorption, the average OCV was lower, with values of about
0.29~V for pristine HfSSe and 0.30~V for Co-doped HfSSe. These voltage
values lay within the desirable low-voltage range for anode materials,
typically around 0.1--1.0~V. Such low OCV values were beneficial because
they could help maintain a high full-cell voltage while remaining
sufficiently above zero to reduce the possibility of metallic Li or Na
plating and dendrite formation. Overall, the specific-capacity and OCV
results demonstrated that Co doping significantly improved the ion-storage
capability of Janus HfSSe without producing any unfavorable voltage
response. The Co-doped HfSSe monolayer exhibited higher theoretical
capacities for both Li and Na storage while maintaining suitable low
average OCV values, indicating its strong potential as an anode material
for Li-ion and Na-ion batteries.


\begin{table*}[t]
\centering
\caption{Calculated diffusion barriers of Li and Na on the S- and
Se-terminated surfaces of pristine and Co-doped HfSSe monolayers.
All values are given in eV.}
\label{tab:diffusion_barriers}
\begin{tabular}{llcccccc}
\hline
\textbf{System} & \textbf{Surface} &
\multicolumn{3}{c}{\textbf{Li diffusion barrier}} &
\multicolumn{3}{c}{\textbf{Na diffusion barrier}} \\
\cline{3-5}\cline{6-8}
 & & Path 1 & Path 2 & Path 3 & Path 1 & Path 2 & Path 3 \\
\hline
Pristine HfSSe  & S side  & 0.2641 & 0.2779 & --     & 0.1355 & 0.1625 & -- \\
Pristine HfSSe  & Se side & 0.2088 & 0.2278 & --     & 0.1173 & 0.1403 & -- \\
Co-doped HfSSe  & S side  & 0.2579 & 0.2473 & 0.4334 & 0.1456 & 0.1542 & 0.2484 \\
Co-doped HfSSe  & Se side & 0.2153 & 0.2279 & 0.2231 & 0.1044 & 0.4118 & 0.1706 \\
\hline
\end{tabular}
\end{table*}

\subsection{Ion diffusion pathways and migration barriers}

The rate capability of an electrode material is strongly influenced by
surface-ion mobility. Therefore, Li and Na diffusion on pristine and
Co-doped Janus HfSSe was investigated using the climbing-image nudged
elastic band method. The S- and Se-terminated surfaces were
considered separately because of the broken out-of-plane symmetry of the
Janus monolayer. The diffusion barrier was determined from the maximum
energy along the minimum-energy pathway relative to the initial state;
lower barriers indicate more facile ion migration.

\begin{figure*}[t]
    \centering
    \includegraphics[width=\textwidth]{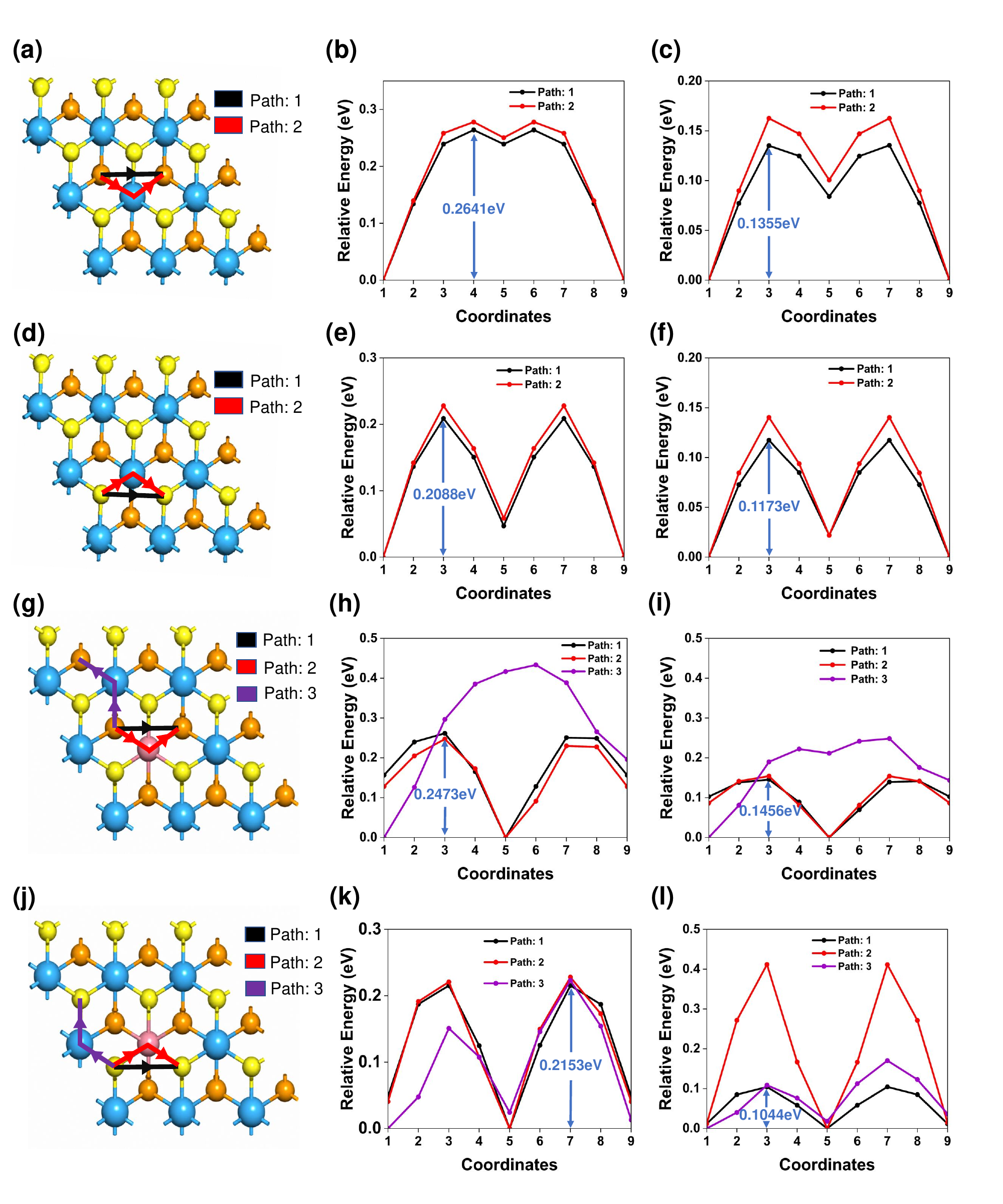}
    \caption{Ion-diffusion pathways and corresponding relative-energy
    profiles on pristine and Co-doped Janus HfSSe monolayers. The first
    column shows the diffusion pathways on (a) the S-terminated surface
    of pristine HfSSe, (d) the Se-terminated surface of pristine HfSSe,
    (g) the S-terminated surface of Co-doped HfSSe, and (j) the
    Se-terminated surface of Co-doped HfSSe. The second column presents
    the corresponding Li diffusion-energy profiles in panels (b), (e),
    (h), and (k), while the third column presents the corresponding Na
    diffusion-energy profiles in panels (c), (f), (i), and (l).}
    \label{fig:diffusion}
\end{figure*}

For pristine HfSSe, two migration pathways were considered on each
surface, as shown in Figs.~\ref{fig:diffusion}(a,d). On the S-terminated
surface, path~1 represents direct
$\mathrm{Se}\rightarrow\mathrm{Se}$ migration, whereas path~2 follows
$\mathrm{Se}\rightarrow\mathrm{Hf}\rightarrow\mathrm{Se}$. On the
Se-terminated surface, the corresponding pathways are
$\mathrm{S}\rightarrow\mathrm{S}$ and
$\mathrm{S}\rightarrow\mathrm{Hf}\rightarrow\mathrm{S}$.
The calculated barriers are summarized in
Table~\ref{tab:diffusion_barriers}. For Li, the minimum barrier was
0.2088~eV along the direct $\mathrm{S}\rightarrow\mathrm{S}$ pathway on
the Se-terminated surface, compared with 0.2641~eV on the S-terminated
surface. Na exhibited still lower barriers, with a minimum value of
0.1173~eV on the Se-terminated surface. Thus, the Se-terminated surface
provided the most favorable diffusion pathway for both Li and Na, while
Na migrated more readily than Li. An additional Hf-to-Hf Li migration
pathway was also examined (see
Supplementary Material).

Co substitution introduced an additional migration pathway through the
dopant site, as shown in Figs.~\ref{fig:diffusion}(g,j). On the
S-terminated surface, paths~1--3 correspond to
$\mathrm{Se}\rightarrow\mathrm{Se}$,
$\mathrm{Se}\rightarrow\mathrm{Co}\rightarrow\mathrm{Se}$, and
$\mathrm{Se}\rightarrow\mathrm{Hf}\rightarrow\mathrm{Se}$,
respectively. On the Se-terminated surface, the corresponding routes are
$\mathrm{S}\rightarrow\mathrm{S}$,
$\mathrm{S}\rightarrow\mathrm{Co}\rightarrow\mathrm{S}$, and
$\mathrm{S}\rightarrow\mathrm{Hf}\rightarrow\mathrm{S}$.
For Li, the lowest barrier on the Co-doped S-terminated surface was
0.2473~eV along the Co-mediated path, whereas the minimum barrier on the
Se-terminated surface was 0.2153~eV along the direct
$\mathrm{S}\rightarrow\mathrm{S}$ route. The latter remained close to
the pristine minimum of 0.2088~eV, indicating that Co substitution
preserved favorable Li mobility.

For Na, the minimum barrier decreased from 0.1173~eV in pristine HfSSe
to 0.1044~eV after Co doping, again along the direct pathway on the
Se-terminated surface. In contrast, the Co-mediated Na pathway on this
surface exhibited a much larger barrier of 0.4118~eV, indicating that
migration directly through the Co-associated region was unfavorable.
Overall, the diffusion behavior remained strongly surface- and
pathway-dependent. Co substitution preserved the low Li migration barrier
while slightly improving the most favorable Na diffusion pathway, and the
consistently lower Na barriers suggested faster Na-ion transport than
Li-ion transport.
\begin{table}[t]
\centering
\caption{Comparison of Li-ion storage performance with other
two-dimensional anode materials.}
\label{tab:li_performance_comparison}
\renewcommand{\arraystretch}{1.10}
\setlength{\tabcolsep}{3pt}
\footnotesize

\resizebox{\columnwidth}{!}{%
\begin{tabular}{lcccc}
\toprule
\textbf{Structure} &
\textbf{\shortstack{Capacity\\(mAh g$^{-1}$)}} &
\textbf{\shortstack{OCV\\(V)}} &
\textbf{\shortstack{Diffusion Barrier\\(eV)}} &
\textbf{Ref.} \\
\midrule

Janus MoSSe      & 776.50 & 0.62 & 0.2400 & \cite{Shang2018MoSSe} \\
Janus WSSe       & 477.80 & 0.48 & 0.1800 & \cite{ahmad2023wsse} \\
Janus CrSSe      & 348.00 & 1.00 & 0.3000 & \cite{Sahoo2024CrXY} \\
Janus CrSTe      & 254.00 & 0.52 & 0.6000 & \cite{Sahoo2024CrXY} \\
Janus CrSeTe     & 208.00 & 0.60 & 0.1000 & \cite{Sahoo2024CrXY} \\
Janus SVSiP$_2$  & 749.02 & 0.26  & 0.2660 & \cite{Ahmad2026SVSiP2} \\
MoS$_2$/Nb$_2$CO$_2$
                  & 412.40 & 0.91 & 0.3100 & \cite{Zhang2026MoS2Nb2CO2} \\
WS$_2$/graphene  & 588.16 & 0.18 & 0.2700 & \cite{Bijoy2024WS2Graphene} \\
HfS$_2$          & 441.86 & 2.78 & 0.2000 & \cite{nayak2025hfs2} \\
VN$_2$           & 678.80 & 0.81 & 0.2011 & \cite{Dong2021VN2} \\
SiP              & 520.00 & 1.50 & 0.2800 & \cite{Xing2024SiP} \\
Pristine HfSSe   & 555.43 & 0.62 & 0.2088 & This work \\
Co-HfSSe         & 776.19 & 0.62 & 0.2153 & This work \\
\bottomrule
\end{tabular}%
}
\end{table}

\begin{table}[t]
\centering
\caption{Comparison of Na-ion storage performance with other
two-dimensional anode materials.}
\label{tab:na_performance_comparison}
\renewcommand{\arraystretch}{1.10}
\setlength{\tabcolsep}{3pt}
\footnotesize

\resizebox{\columnwidth}{!}{%
\begin{tabular}{lcccc}
\textbf{Structure} &
\textbf{\shortstack{Capacity\\(mAh g$^{-1}$)}} &
\textbf{\shortstack{OCV\\(V)}} &
\textbf{\shortstack{Diffusion Barrier\\(eV)}} &
\textbf{Ref.} \\
\midrule

Janus HfSTe      & 554.80 & 0.33 & 0.1800 & \cite{Pham2026HfSTe} \\
Janus WSSe       & 371.50 & 0.57 & 0.0400 & \cite{ahmad2023wsse} \\
Janus CrSSe      & 260.00 & 0.10 & 0.5400 & \cite{Sahoo2024CrXY} \\
Janus CrSTe      & 198.00 & 0.49 & 0.3200 & \cite{Sahoo2024CrXY} \\
Janus CrSeTe     & 177.00 & 0.51 & 0.1500 & \cite{Sahoo2024CrXY} \\
Janus V$_2$COS   & 662.18 & 0.76 & 0.1200 & \cite{Wang2024V2COS} \\
Janus BeSeCl     & 699.65 & 0.91 & 0.0800 & \cite{Verma2025BeSeCl} \\
Y$_2$B$_2$       & 403.16 & 0.45 & 0.0080 & \cite{Gao2021Y2B2} \\
Borophosphene    & 1282.00 & 0.367 & 0.2210 & \cite{Zhang2020Borophosphene} \\
BC$_3$N$_2$      & 777.00 & 0.72 & 0.1500 & \cite{Ye2024BC3N2} \\
Pristine HfSSe   & 370.29 & 0.29 & 0.1173 & This work \\
Co-HfSSe         & 582.14 & 0.30 & 0.1044 & This work \\
\bottomrule
\end{tabular}%
}
\end{table}

\section{Comparative analysis}

To evaluate the relative performance of pristine and Co-doped HfSSe as
anode materials, representative two-dimensional anodes are compared in
Tables~\ref{tab:li_performance_comparison} and
\ref{tab:na_performance_comparison} for Li-ion and Na-ion storage,
respectively. For Li-ion storage, Co-HfSSe delivered a theoretical
capacity of 776.19~mAh~g$^{-1}$, approximately twice the theoretical
capacity of commercial graphite~\cite{Grey2023Graphite}, which is
372~mAh~g$^{-1}$. This capacity
exceeded those of other Janus monolayers such as
WSSe~\cite{ahmad2023wsse}, CrSSe, CrSTe, CrSeTe~\cite{Sahoo2024CrXY},
and SVSiP$_2$~\cite{Ahmad2026SVSiP2}, and was comparable to the
776.50~mAh~g$^{-1}$ reported for Janus MoSSe~\cite{Shang2018MoSSe}.
With an OCV of 0.62~V, similar to that of MoSSe, Co-HfSSe also showed a
lower Li migration barrier (0.2153~eV) than MoSSe
(0.24~eV)~\cite{Shang2018MoSSe}, WS$_2$/graphene~\cite{Bijoy2024WS2Graphene},
and SiP~\cite{Xing2024SiP}. Compared with the related Hf-based
dichalcogenide HfS$_2$, which exhibited a capacity of
441.86~mAh~g$^{-1}$ and a high OCV of 2.78~V~\cite{nayak2025hfs2},
Co-HfSSe offered a substantially higher capacity at a much lower
operating voltage, which is more suitable for achieving a high full-cell
voltage. It also outperformed VN$_2$~\cite{Dong2021VN2} and the
MoS$_2$/Nb$_2$CO$_2$ heterostructure~\cite{Zhang2026MoS2Nb2CO2} in
capacity while maintaining a lower OCV. For Na-ion storage, Co-HfSSe
exhibited a capacity of 582.14~mAh~g$^{-1}$, higher than those of Janus
HfSTe~\cite{Pham2026HfSTe}, WSSe~\cite{ahmad2023wsse}, and
Y$_2$B$_2$~\cite{Gao2021Y2B2}, together with a low OCV of 0.30~V and a
small diffusion barrier of 0.1044~eV. Although Janus
V$_2$COS~\cite{Wang2024V2COS} and BeSeCl~\cite{Verma2025BeSeCl} showed
slightly higher Na capacities and comparable diffusion barriers, their
OCVs (0.76 and 0.91~V, respectively) were considerably higher than that
of Co-HfSSe. Similarly, borophosphene~\cite{Zhang2020Borophosphene} and
BC$_3$N$_2$~\cite{Ye2024BC3N2} offered higher Na capacities but at
higher operating voltages and larger diffusion barriers. Overall,
Co-HfSSe provided a balanced combination of high capacity, low operating
voltage, and fast ion migration for both Li-ion and Na-ion storage.

\section{Conclusion}

In this work, pristine and Co-doped Janus HfSSe monolayers were systematically investigated as anode materials for Li-ion and Na-ion batteries using first-principles calculations. The optimized pristine monolayer was found to be dynamically stable through phonon calculations, while the AIMD simulation confirmed that the Li-adsorbed Co-doped structure remained thermally stable at 300~K. Pristine HfSSe exhibited an indirect band gap of approximately 0.85~eV, whereas Co substitution shifted the Fermi level into the valence-band region and introduced weak spin polarization. More importantly, adsorption of the first Li or Na layer produced bands crossing the Fermi level and a finite density of states at $E_{\mathrm{F}}$, indicating a transition to metallic behavior. This adsorption-induced metallization was expected to facilitate electronic transport during battery operation. Both Li and Na interacted favorably with the two chemically distinct surfaces of Janus HfSSe. Co substitution substantially strengthened these interactions by creating an active adsorption center and modifying the local electronic environment. The most favorable adsorption energies changed from $-2.6125$ to $-3.7405$~eV for Li and from $-1.9800$ to $-3.0529$~eV for Na after Co doping. Bader charge analysis revealed that the substituted Co donated only
$\sim$0.58~$e$ compared with $\sim$2.1~$e$ for Hf, leaving the
neighboring chalcogen atoms electron-deficient; these empty
chalcogen-$p$ states accepted the electron donated by the adsorbed
alkali atom at lower energy, which explained the stronger binding
despite a nearly unchanged charge transfer of $\sim$0.88~$e$. Co doping also significantly enhanced the multilayer storage capability of HfSSe. The maximum Li loading increased from three layers in pristine HfSSe to four layers in the Co-doped structure, resulting in an increase in theoretical capacity from 555.43 to 776.19~mAh~g$^{-1}$. Similarly, the maximum Na loading increased from two layers to three layers, raising the capacity from 370.29 to 582.14~mAh~g$^{-1}$. These improvements were achieved while maintaining favorable average open-circuit voltages of approximately 0.62~V for Li and 0.29--0.30~V for Na. The calculated minimum diffusion barriers remained low, with values of 0.2088 and 0.2153~eV for Li on pristine and Co-doped HfSSe, respectively. For Na, Co doping lowered the minimum barrier from 0.1173 to 0.1044~eV, indicating particularly rapid Na-ion transport along the preferred pathway. Comparison with representative two-dimensional anode materials demonstrated that Co-doped HfSSe provided a competitive balance among storage capacity, operating voltage, electronic conductivity, and ion mobility. Overall, the results identified Co substitution as an effective strategy for improving the electrochemical performance of Janus HfSSe and established Co-doped HfSSe as a promising anode candidate for both Li-ion and Na-ion batteries.


\appendix


\printcredits
\section*{Declaration of competing interest}

The authors declare that they have no known competing financial interests
or personal relationships that could have appeared to influence the work
reported in this paper.

\section*{Data availability}

The data supporting the findings of this study are available upon request.

\section*{Acknowledgments}

The authors gratefully acknowledge the support provided by the Department of Electrical and Electronic Engineering, Bangladesh University of Engineering and Technology (BUET). The authors also thank the Institute of Information and Communication Technology (IICT), BUET, for providing the computational facilities and resources used in this work.

\appendix
\section{Supplementary data}

Supplementary material related to this article is
\href{https://github.com/ahmedzubair003/HfSSe_Batteries}{available here}.

\bibliographystyle{model1-num-names}
\bibliography{cas-refs}

\end{document}